\documentclass[prb,superscriptaddress,twocolumn,showpacs,amsmath,amssymb]{revtex4}

\usepackage{graphicx}
\usepackage{dcolumn}
\usepackage{bm}

\begin{document}

\title{Extracting the spectral function of the cuprates by a full two-dimensional analysis: Angle-resolved photoemission spectra of Bi$_{2}$Sr$_{2}$CuO$_{6}$ }

\author{W. Meevasana}
\email[]{non@stanford.edu}
\affiliation {Department of Physics,
Applied Physics, and Stanford Synchrotron Radiation Laboratory,
Stanford University, Stanford, CA 94305}

\author{F. Baumberger}
\affiliation {School of Physics and Astronomy, University of St.
Andrews, St. Andrews, Fife KY16 9SS, UK}

\author{K. Tanaka}
\affiliation {Department of Physics, Applied Physics, and Stanford
Synchrotron Radiation Laboratory, Stanford University, Stanford,
CA 94305} \affiliation {Advanced Light Source, Lawrence Berkeley
National Lab, Berkeley, CA 94720}

\author{F. Schmitt}
\affiliation {Department of Physics, Applied Physics, and Stanford
Synchrotron Radiation Laboratory, Stanford University, Stanford,
CA 94305}

\author{W. R. Dunkel}
\affiliation {Department of Physics, Applied Physics, and Stanford
Synchrotron Radiation Laboratory, Stanford University, Stanford,
CA 94305}

\author{D. H. Lu}
\affiliation {Department of Physics, Applied Physics, and Stanford
Synchrotron Radiation Laboratory, Stanford University, Stanford,
CA 94305}

\author{S.-K. Mo}
\affiliation {Department of Physics, Applied Physics, and Stanford
Synchrotron Radiation Laboratory, Stanford University, Stanford,
CA 94305} \affiliation {Advanced Light Source, Lawrence Berkeley
National Lab, Berkeley, CA 94720}

\author{H. Eisaki}
\affiliation{Nanoelectronic Research Institute, AIST, Tsukuba
305-0032, Japan}

\author{Z.-X. Shen}
\email[]{zxshen@stanford.edu}
\affiliation {Department of Physics,
Applied Physics, and Stanford Synchrotron Radiation Laboratory,
Stanford University, Stanford, CA 94305}

\date{\today}

\begin{abstract}
Recently, angle-resolved photoemission spectroscopy (ARPES) has
revealed a dispersion anomaly at high binding energy near 0.3-0.5
eV in various families of the high-temperature superconductors.
For further studies of this anomaly we present a new
two--dimensional fitting-scheme and apply it to high-statistics
ARPES data of the strongly-overdoped Bi$_{2}$Sr$_{2}$CuO$_{6}$
cuprate superconductor. The procedure allows us to extract the
self-energy in an extended energy and momentum range. It is found
that the spectral function of Bi$_{2}$Sr$_{2}$CuO$_{6}$ can be
parameterized using a small set of tight-binding parameters and a
weakly-momentum-dependent self-energy up to 0.7 eV in binding
energy and over the entire first Brillouin zone. Moreover the
analysis gives an estimate of the momentum dependence of the
matrix element, a quantity, which is often neglected in ARPES
analyses.
\end{abstract}

\maketitle

\section{\label{introp}Introduction}

Angle-resolved photoemission spectroscopy (ARPES) has been an
excellent tool for studying many-body interactions in
two-dimensional strongly-correlated systems \cite{Group:Review}.
Recently, ARPES studies revealed a new energy scale in the form of
a large dispersion anomaly near 0.3-0.5 eV below $E_F$, seen in
various families of cuprates, at a wide range of doping and
measuring conditions (e.g. photon energy) \cite{CCOC:Filip,
HEA:Non, HEA:Lanzara, HEA:Donglai, HEA:Valla, HEA:Pan, HEA:Chang}.
Given its phenomenological behavior, this anomaly could be
important to understand the nature of high-temperature
superconductivity. However, its origin is still under debate
\cite{CCOC:Filip, HEA:Non, HEA:Lanzara, HEA:Donglai, HEA:Valla,
HEA:Pan, HEA:Chang, Hubbard:Hanke, Hubbard:Alexandru,
tJ:Manousakis, DMFT:Byczuk, Magnon:Markiewicz, crossover:Wang,
InplaneOptical:Chubukov}.

From earlier study, this energy scale is in the range where the
\emph{J} scale coherent band is split from the \emph{t} scale
incoherent electronic structure structure due to Mott-Hubbard
physics \cite{Review:Dagotto}. Newer data seem to suggest that
there is momentum dependence even in the incoherent part of the
electronic state at higher energy, even though the uncertainty due
to matrix element distortion has been raised \cite{ME:Inosov,
ME:Kordyuk}. An interesting question would then be whether one can
take the data in its face value and make a global analysis in
terms of self-energy. This has several advantages. First, it
explores a new ARPES methodology to extract and parameterize many
body effects in complex materials. Second, it allows one to make a
comparison with the information extracted from optical
reflectivity measurements \cite{InplaneOptical:vanderMarel,
InplaneOptical:Chubukov, InplaneOptical:Erik,
InplaneOptical:Hwang} where one would not expect the same kind of
matrix element effects as in ARPES. This would be a good
consistency check. Last, it will help us to gain insights of
many-body effect and matrix element effect even in the ARPES
context. Such an approach is non-trivial to implement due to the
difficulties encountered with standard data analysis techniques,
which can introduce strong artifacts in the extracted quantities
of interest, most notably the self-energy. An improved
quantitative analysis of the experimental data would undoubtedly
help to advance our understanding.

It is a technical challenge to extract the spectral function,
which contains the information of the interactions, from ARPES
data. The main problem is the lack of general analytic expressions
for individual momentum distribution curves (MDCs) or energy
distribution curves (EDCs). Non-Lorentzian MDC peak shapes are
common even in simple situations, e.g. in the case of non-linearly
dispersing bands. Energy distribution curves are even more
delicate to analyze since their precise shape is determined by the
energy dependence of the self-energy, i.e. the quantity that
should be extracted from the analysis. Further complications arise
from the finite instrumental energy and momentum resolution,
low-counting statistics, or matrix element effects. All these
complications cause discrepancies between the commonly used MDC or
EDC analyzes, which become, particularly pronounced at high
binding energy \cite{HEA:Lanzara, HEA:Donglai}. Different problems
arise in the other important regime near the Fermi level where the
feature of interest (e.g. the scattering rate near the Fermi
level) are comparable in width to the instrumental resolution. In
this case, one needs to estimate the combined effect of energy and
momentum resolution on a single EDC or MDC, which can only be done
approximately or by restricting the self-energy to simple analytic
forms \cite{RhO4:FelixB, RuO4:Nik}.

In this paper we introduce a new two-dimensional (2D) analysis
scheme, which allows us to extract an empirical spectral function
of a strongly interacting system over an extended energy range.
The method is applied to high-statistics ARPES data from
strongly-overdoped Bi$_{2}$Sr$_{2}$CuO$_{6}$ (Bi2201) single
crystals. The feature of interest will be the high-energy anomaly
around 0.3-0.5 eV. We will not concern ourselves with the low
energy anomaly or "kink" (0.03-0.09eV), which can be highly
momentum dependent \cite{Bi2212:Tanja}. Since the width of the
high-energy anomaly is large compared to the experimental
resolution, we will neglect the influence of instrumental
broadening. Our analysis qualitatively reproduces all the basic
features seen in both MDC and EDC analysis and shows that the
spectral function of Bi2201 can be empirically parameterized by a
simple and compact set of tight-binding parameters fitted to
local-density-approximation (LDA) calculations and a weakly
momentum-dependent self-energy. Further, this extracted
self-energy is in reasonable agreement with the one extracted from
optical reflectivity measurement \cite{InplaneOptical:Erik}. This
finding provides a new approach to understanding many-body effects
beyond the narrow energy range around the Fermi level which has
traditionally been the focus of ARPES studies.

We have selected strongly-overdoped Pb-substituted Bi2201 for this
study for several reasons: (a) in the overdoped regime, there is
no complication from pseudogap behavior near the antinodal region
or from polaronic behavior; hence the 2D analysis can be applied
to the whole Brillouin Zone (BZ); (b) bi-layer splitting effects
are absent in this single-layer cuprate and superlattice effects
are largely suppressed by the Pb content; (c) measurements at low
temperature, where thermal broadening is small, are not
complicated by the effects of the superconducting gap.

\section{\label{resigmap}Experiment}

We have measured single crystals of Pb-substituted Bi2201. The
overdoped (OD) samples with composition,
Pb$_{0.38}$Bi$_{1.74}$Sr$_{1.88}$CuO$_{6+\delta}$, are
non-superconducting ($\rm T_c$ $<$ 4 K). ARPES data were collected
on a Scienta R4000 electron energy analyzer at the Advanced Light
Source (ALS) with photon energies of 42 and 55 eV and a base
pressure of $4 \times 10^{-11}$ torr. This analyzer has the
advantage of a large-angle window which can cover the band
dispersion across the entire BZ. Samples were cleaved \emph{in
situ} in the normal state at the measurement temperature of 20K.
The energy resolution was set to 13-18 meV. The average momentum
resolution at these photon energies was ~ 0.021 {\AA}$^{-1}$ (or
0.35$^{\circ}$). The linear polarization of the light source is
fixed to be in-plane along (0,0) to ($\pi$,$\pi$) through out the
measurement. Note that the fitted matrix element, which will be
shown in the following, should be referred to this particular
experimental geometry.

\section{\label{resigmap}2D analysis}
Conventionally, ARPES data from cuprates are analyzed by fitting
large numbers of one-dimensional intensity profiles at constant
energy (MDCs) or constant momentum (EDCs) using simple analytical
functions. However, such an analysis of EDCs or MDCs has
limitations which will be discussed in more detail in Appendix A
(Fig. 5). Attempting to go beyond the conventional EDC or MDC
analysis , we use here a full 2D analysis which will be explained
in the following. Our starting point is the common expression for
the photocurrent within the sudden approximation
\cite{Group:Review}:
\begin{equation}
I({\bf k},\omega) =I_0({\bf k},\nu,{\bf A}) f(\omega){\mathcal
A}({\bf k},\omega)
\end{equation}
where $I_0({\bf k},\nu,{\bf A})$ is proportional to the squared
one-electron matrix element and depends on in-plane electron
momentum ${\bf k}$, the energy ($\nu$) and polarization (or vector
potential, ${\bf A}$) of the incoming photon, $f(\omega)$ is the
Fermi function. $\mathcal A({\bf k},\omega)$ is the
single-particle spectral function that contains all the
corrections from the many-body interactions in the form of the
self-energy, $\Sigma({\bf k}, \omega)$,
\begin{equation}
{\mathcal A}({\bf k},\omega)  = \frac{(-1/\pi)\ \ \textrm{Im}
\Sigma({\bf k},\omega)}{[\omega -
\epsilon^{0}_{k}-\textrm{Re}\Sigma({\bf
k},\omega)]^{2}+[\textrm{Im}\Sigma({\bf k},\omega)]^{2}}
\end{equation}
where $\epsilon^{0}_{k}$ is the bare band dispersion. Note that we
have neglected the instrumental resolution in Eq. 1.

This form is intrinsically multi-dimensional. Given that ARPES
data are collected with most modern spectrometers in parallel as a
function of energy and one momentum coordinate, it appears to be
an artificial oversimplification to analyze the data by fitting
single line profiles. Instead, the analysis presented below is an
attempt to fit ARPES data at once in 2D images. This analysis
assumes a simple form of the bare band dispersion given by a
tight-binding approximation of LDA calculations \cite{HEA:Non}.
For simplicity , we will also assume weak momentum dependence of
the self-energy (i.e. over a sufficiently small k space range,
$\Sigma({\bf k},\omega)$ $\rightarrow$ $\Sigma (\omega)$) where in
the next section, we will show that this is a reasonable
assumption. However, our analysis does not assume any particular
form of the self-energy and the matrix element. This is achieved
by assigning an individual fit parameter for real and imaginary
part of the self-energy to every measured energy point and a fit
parameter for the matrix element to every momentum point.

\begin{figure}[t]
\includegraphics [width=3.2in, clip]{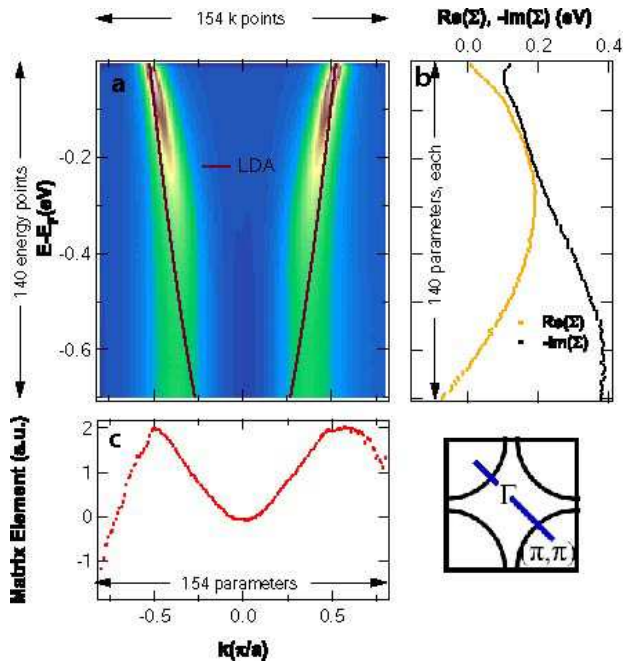}
\caption{\label{fig1} (Color online) a) Image plot of the fitted
intensity from the 2D analysis of ARPES data of OD Bi2201 along
the nodal direction (0,0) to ($\pi, \pi$) where the solid line
shows a bare dispersion derived from LDA. b) The fitting
parameters for real and imaginary parts of self-energy. c) The
fitting parameters for the matrix element. }
\end{figure}

This is illustrated in Fig. 1 showing an image plot of the fitted
intensity together with the values of the fitting parameters for
the self-energy ($\Sigma$) and matrix element ($I_0({\bf
k},\nu,{\bf A})$). The bare dispersion $\epsilon^{0}_{k}$  derived
from LDA is shown as solid line. Starting with an ARPES image of
154 by 140 data points in momentum and energy respectively, the
least-square fit of this Fig. 1 includes 154 parameters for each k
point for the matrix element, 140 parameters for each energy point
of $\textrm{Im}\Sigma$ and $\textrm{Re}\Sigma$, and a few
additional parameters for an overall intensity and background,
i.e. a total of about 440 parameters. This number seems high, but
it is justifiable given that the number of data points is much
larger. In the above example, we have $140\times154$ = 21560 data
points corresponding to about 50 points per parameter. Fitting a
single MDC with a Lorentzian on a constant background requires
more parameters per data-point.

We stress again that this 2D analysis on Bi2201 will focus on the
high-energy anomaly whose energy scale of 0.3-0.5eV is larger than
the energy resolution. We will not focus on the energy scales
below 0.1 eV where it has been shown that the self-energy is
strongly momentum dependent \cite{Bi2212:Tanja}. We also note that
Kramers-Kronig consistency of the self-energy is not implemented
in the fitting procedure. This will be discussed in more detail in
Appendix B.

\section{\label{resigmap}Results}

\subsection{\label{Akw} Extracting Spectral Function and Self-energy}

\begin{figure*}
\begin{center}
\includegraphics [width=5.5in, clip]{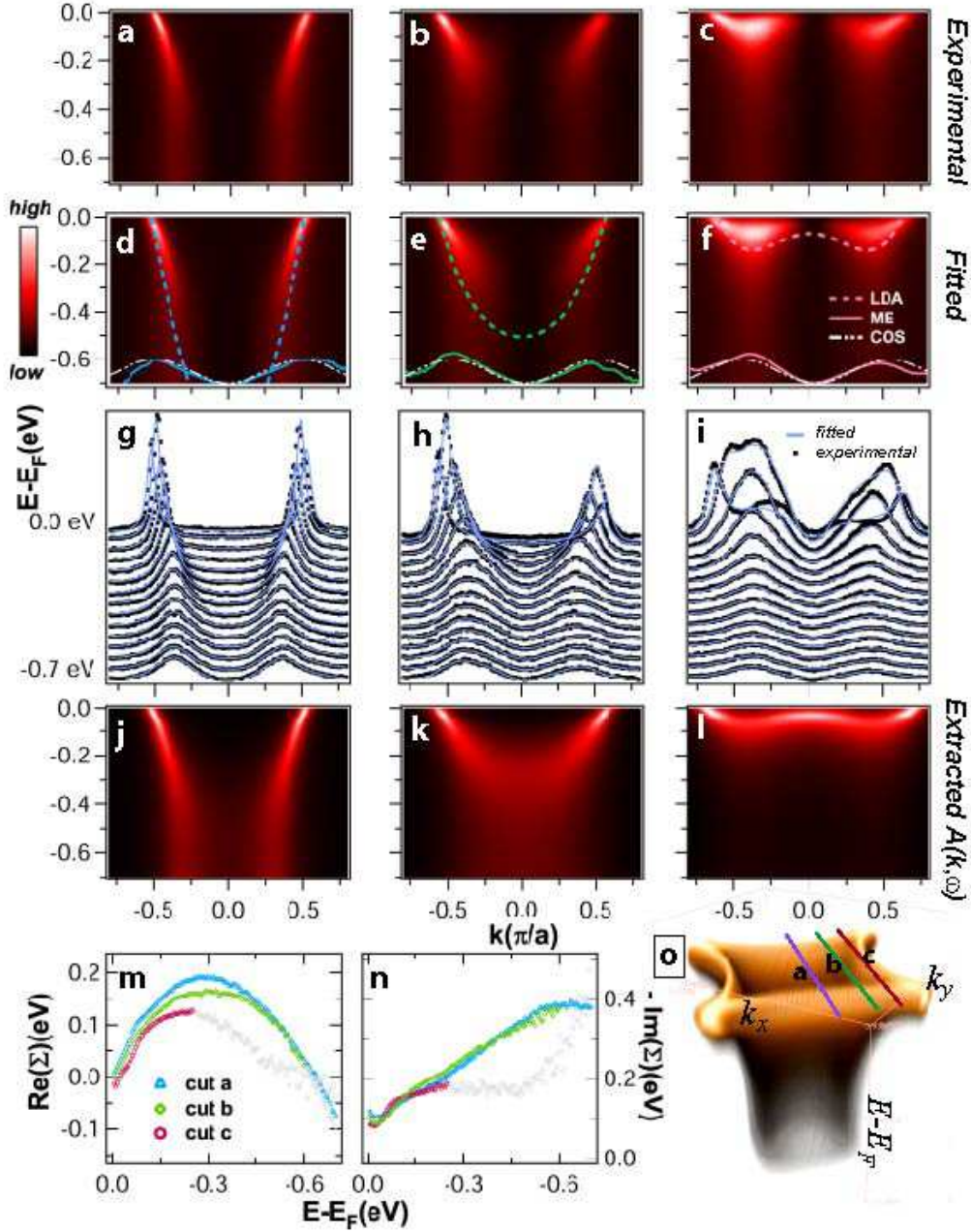}
\end{center}
\caption{\label{fig2} (Color online) On the first row, a)-c) are
the experimental ARPES data along momentum direction as indicated
by the band shown in bottom right. On the second row, d)-f) are
the corresponding image plots of the fitted intensity from the 2D
analysis. The thicker color dashed lines are the dispersions
generated from the tight-binding parameters from LDA calculation
(LDA) given above. The solid color lines are the matrix elements
(ME) obtained from the 2D analysis where the smaller white dashed
lines are the empirically guessed form of the matrix element in
the form of a cosine function (COS). On the third row, g)-i) are
the MDCs of raw data (black dots) and corresponding image plot of
the fitted intensity (blue line). Taken out the matrix element
effect and background, j)-l) are the corresponding extracted
spectral function ($\mathcal A({\bf k},\omega)$) from the 2D
analysis. m) and n) are the extracted real and imaginary parts of
the self-energy in Eq. 2 for the cuts a, b and c; the extracted
values are plotted in colors up to the energy not far from the
bottom of the bare band (up to 0.6 eV for cut b and 0.25 eV for
cut c) and in grey at higher energy. o) shows the band structure
generated from the tight-binding parameters and the self-energy
along the nodal direction.}
\end{figure*}

We now apply the 2D analysis to several k-space cuts as shown in
Fig. 2. The ARPES experimental data are shown in the first row
(Fig. 2(a)-2(c)), taken with photon energy 42 eV. We deliberately
choose the spectra which would be difficult to analyze by MDC or
EDC analysis alone because they have regions where EDC (see, cut
a) and MDC (see, cut c) peaks are not be well-defined. We then
apply the 2D analysis on these data using the following tight
binding description of the bare dispersion: $E(k) = -2t[cos(k_x
a)+ cos(k_y b)]-4 t' cos(k_x a) cos(k_y b) - 2t''[cos(2 k_x a)+
cos(2 k_y b)] - E_F$ where $t = 0.435, t'= -0.1, t''= 0.038,$ and
$E_F = -0.5231$ eV\cite{HEA:Non, TB:comment}. Since the background
is small ($\sim$5\% of the spectral intensity at $k_F$) and flat
in this chosen energy range (EDCs are shown in Fig. 1(c) of Ref.
\cite{HEA:Non}), we use a constant energy-and-momentum-independent
background for this particular data.

The corresponding fits are very well in agreement as shown in the
second row (Fig. 2(d)-2(f)) while MDCs of the fit and the raw data
are shown in the third row (Fig. 2(g)-2(i)). From this 2D
analysis, we then extract the spectral function ($\mathcal A({\bf
k},\omega)$) which is shown in Fig. 2(j)-2(l). These plots of
extracted spectral function do not contain the matrix elements
anymore but still show the high-energy dispersion anomaly.
Further, the optics matrix element effect is different and much
weaker and an extracted self-energy of the same Bi2201 sample
obtained from optical reflectivity measurement shows a reasonable
agreement\cite{InplaneOptical:Erik}. The above analysis as well as
the consistency with optics suggests that there is a real many
body anomaly in the energy range. We should note that although we
agree that matrix element effects may distort the spectral line
shape (e.g. the difference between Fig. 2(a)-2(c) and 2(j)-2(l),
respectively), they can hardly explain the high-energy anomaly as
put forward by Ref. \cite{ME:Inosov}. Possibly, the effects
observed and discussed there are influenced in a non-negligible
way by multi-band effects known in Bi2212 and YBCO.

The self-energies ($\Sigma$) extracted by the 2D analysis in Fig.
2(m) and 2(n) show only weak momentum dependence over the most
relevant energy range. Pronounced differences between the three
cuts shown in Fig. 2 are only found at energies below the band
bottom. However, these parts of the self-energies (see grey
symbols in Fig. 2(m) and 2(n)) should not be overestimated as they
will not contribute much spectral weight to the spectra. Hence,
the ARPES data of this OD Bi2201 system can approximately be
parameterized in a simple form using tight-binding parameters and
the self-energy along the nodal direction without losing much
information of the spectral weight distribution throughout energy
and momentum space. Therefore, we can calculated a good
approximation of the full 3D intensity distribution from this
information as shown in Fig. 2(o).

This is intriguing finding that the information of the ARPES
spectra of the OD Bi2201 system \emph{throughout the BZ} can be
much deduced into a very simple and compact set of tight-binding
parameters and a weakly-momentum dependent self-energy. Given its
simplicity, we believe that this finding will reveal us more of
the nature of the interactions in cuprates, especially of the
high-energy anomaly \cite{CCOC:Filip, HEA:Non, HEA:Lanzara,
HEA:Donglai, HEA:Valla, InplaneOptical:Chubukov}.

\begin{figure}[t]
\includegraphics [width=2.9in, clip]{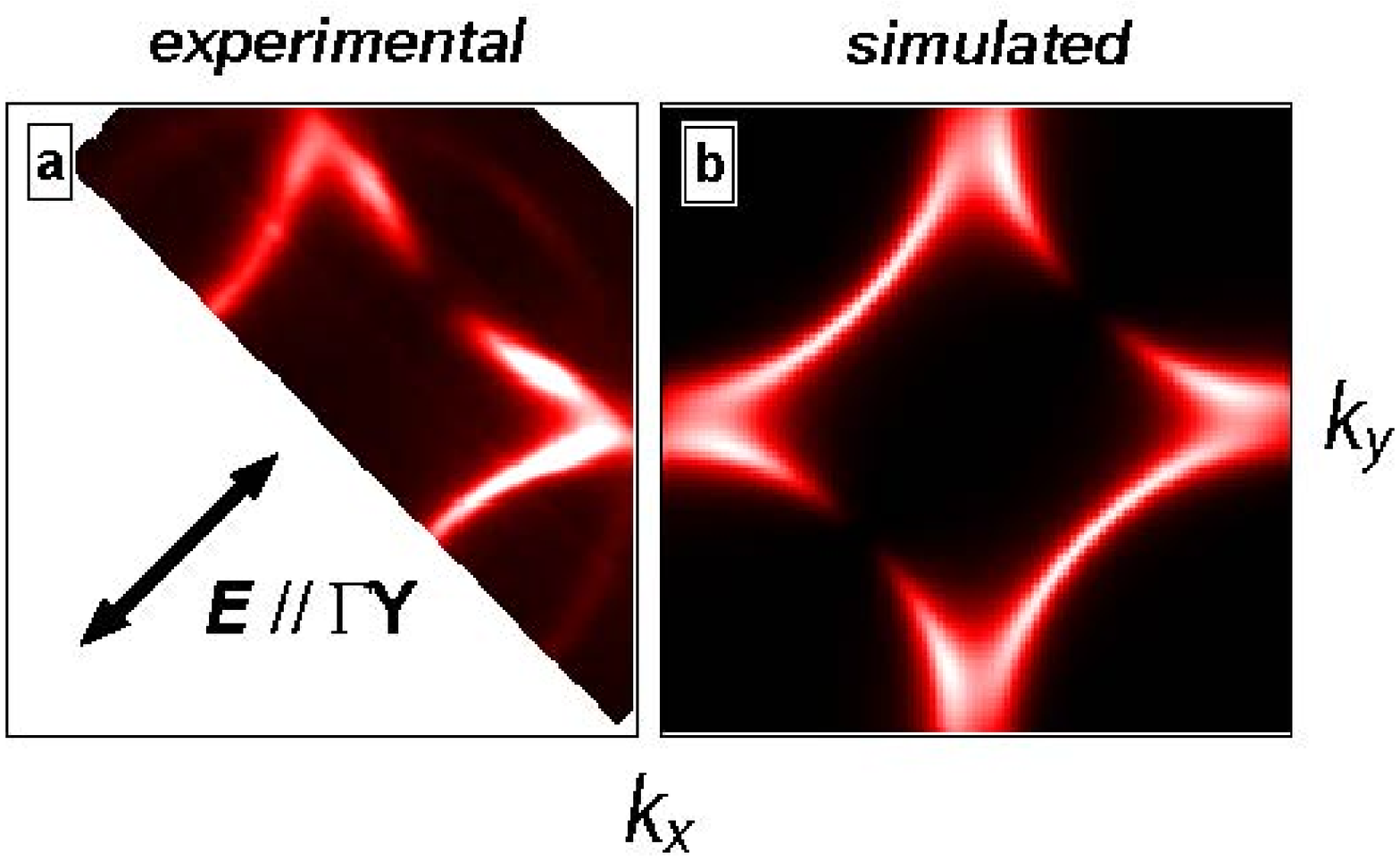}
\caption{\label{fig3} (Color online) a) shows the experimental
Fermi surface map of OD Bi2201 where the light polarization, E, is
along the nodal direction, $\Gamma \rightarrow Y$. b) shows the
Fermi surface map which is constructed from the extracted spectral
function and the empirical guessed matrix element.}
\end{figure}

\subsection{\label{Matrix} Matrix Element Effect}

Although the matrix element effect is known to exist in ARPES
measurement \cite{ME:Bansil}, matrix elements are often neglected
in the analysis of ARPES data, whereas they are naturally embedded
in the presented 2D analysis. This uniquely allows one to separate
the spectral function from matrix element effects as demonstrated
in Fig. 2(j)-2(l). The behavior of the matrix elements, as
obtained from our analysis is shown in arbitrary units at the
bottom of Fig. 2(d)-2(f) by solid colored lines. The line shape of
the extracted matrix element are usually smooth in the region of
$|k| < k_{F}$ but will get noisy outside the Fermi surface. This
can be explained by the following. In the region outside $k_F$,
there is not much of the spectral weight to be fitted and hence,
we emphasize that only in region of $|k| < k_{F}$, the extracting
of matrix element should be counted. As shown in Fig. 2(d)-2(f),
the extracted matrix elements in the $|k| < k_{F}$ region show
similar line shape. Here, we empirically guess the form of the
matrix element to be in the form of $\alpha (1-\cos(2 \vec{k}\cdot
a\hat{s}))/2$ where $\hat{s}$ is the direction of the light
propagating vector (perpendicular to the polarization, $\hat{E}$)
and $\alpha$ is an arbitrary constant along $\hat{s}$; these
empirical forms are shown as the white dash lines on top of the
extracted matrix elements.

As a cross test of our analysis, we construct the Fermi surface
using the extracted spectral function and the empirical form of
the matrix element with $\alpha = 1$. In Fig. 3, we compare this
constructed Fermi surface of OD Bi2201 (Fig. 3(b)) to the
experimental data (Fig. 3(a)). The one-step model calculation by
Mans \emph{et. al.} for Bi2212 system \cite{shadow:Mans} shows a
similar intensity distribution. Although this empirical form of
the matrix element may be oversimplified, some of main features
(e.g. the suppression of the intensity along $\Gamma\rightarrow
Y$) could already be captured by this form.

\begin{figure}[t]
\includegraphics [width=3.4in, clip]{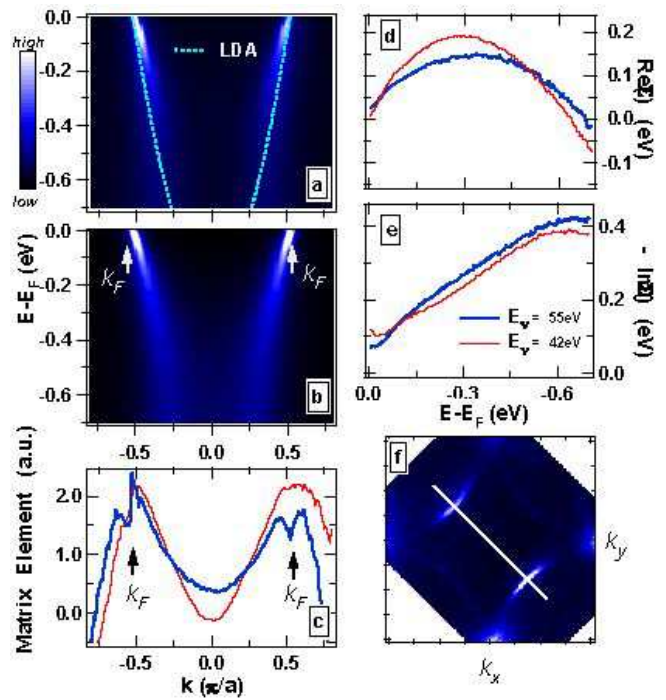}
\caption{\label{fig4} (Color online) Measured at photon energy,
$E_{\nu}$ = 55 eV, a) shows the experimental ARPES data along
(0,0) to ($\pi, \pi$) direction as indicated in f) where the dash
line shows the tight-binding band from LDA calculation used as
bare dispersion. b) shows the corresponding extracted spectral
function ($A(k,\omega)$) from the 2D analysis. c) show the
extracted matrix elements of the data taken at $E_{\nu}$ = 55 eV
(solid line) and 42 eV (dash line) where the areas under the
graphs between $k < |k_F|$ are normalized to be the same. d) and
e) are the extracted $\textrm{Re}\Sigma$ and $\textrm{Im}\Sigma$
of the data taken at $E_{\nu}$ = 55 eV (solid line) and 42 eV
(dash line). f) shows the Fermi surface map.}
\end{figure}

\subsection{\label{55eV} Comparison of Data Measured at Two Different Photon Energies}

To check further on the robustness of the 2D analysis, we perform
ARPES measurement with a second photon energy (55 eV) (Fig. 4),
taken on a different sample and  compare them with the 42 eV data
shown in Fig. 2 and 3. A clear difference in intensity modulation
of Fermi surface maps with these two photon energies is evident
from a comparison of Fig. 3(a) and 4(f). We then apply the 2D
analysis on this 55 eV data. As shown in Fig. 4(c), the extracted
matrix element of 55 eV data (solid line) looks different in
curvature from the 42 eV data as expected from the different
appearance of the two Fermi surface maps. And, as shown in Fig.
4(d) and 4(e), the extracted self-energies of 42 eV and 55 eV show
the same line shape but have slight differences in energy value,
giving the impression of error bar from this 2D analysis. Notice
that the comparison of the extracted $\textrm{Im}\Sigma$ gives
better agreement than $\textrm{Re}\Sigma$. We believe that one
possibility might be due to that $\textrm{Re}\Sigma$ couples
directly to the k-dependent bare band dispersion
$\epsilon^{0}_{k}$ (see Eq. 2) and hence a slight misalignment of
k space from experiment or any $k_z$ dependent effects from
different photon energies \cite{3D:Takeuchi} could cause a larger
error bar to the extracted value of $\textrm{Re}\Sigma$.

\section{\label{con}CONCLUSION}

We presented a 2D analysis method for ARPES data, which allows one
to extract self-energies and matrix elements in more general
situations and with much higher reliability as compared to
standard line-by-line analyses. The method has been applied to
high-statistics ARPES data from strongly overdoped Bi2201. It was
found that the spectral function at high energies is well
approximated over the entire Brillouin zone by a very simple and
compact set of tight-binding parameters and a weakly-momentum
dependent self-energy. We believe that this method will be useful
in the analysis of ARPES data from many systems and may provide us
with more information for improving understanding of many-body
interactions in cuprates.

\begin{acknowledgments}
W.M. would like to thank D. van der Marel, E. van Heumen, T.P.
Devereaux, B. Moritz and N. J. C. Ingle for helpful discussions.
The work at SSRL and ALS are supported by DOE's Office of Basic
Energy Sciences under Contracts No. DE-AC02-76SF00515 and
DE-AC03-76SF00098. This work is also supported by DOE Office of
Science, Division of Materials Science, with contract
DE-FG03-01ER45929-A001 and NSF grant DMR-0604701. W.M.
acknowledges DPST for the financial support.
\end{acknowledgments}

\appendix

\section{MDC and EDC analaysis}

\begin{figure}[t]
\includegraphics [width=2.1in, clip]{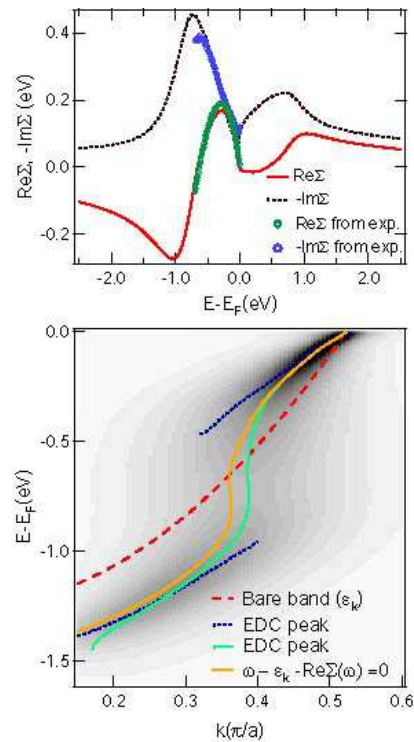}
\caption{\label{fig5} (Color online) a) shows the real (solid
line) and imaginary (dash line) parts of self-energy used to
construct the spectral image shown in b). In Fig. b), we compare
the MDC-peak (fine dash line) and EDC-peak (medium-size dash line)
dispersions with the input band dispersion (solid line). We note
that $\textrm{Re}\Sigma$ and $\textrm{Im}\Sigma$ shown in a)
satisfy the Kramers-Kronig relation where the diamond and circle
symbols represent extracted real and imaginary parts of
self-energy, respectively from Fig. 2(m) and 2(n) of cut a.}
\end{figure}

MDC and EDC analysis is the conventional method to analyze each
\emph{one dimensional} (1D) spectrum of ARPES data. For an
example, a peak position of an EDC could represent the band
dispersion and its peak width could represent the scattering rate.
However, a common problem occurring with these analysis is that a
good fitting of an experimental data set cannot be done
empirically but often needs an additional theoretical modeling. To
get a precise fitting of an EDC, or an 1D \emph{energy}-dependent
spectrum at a fixed momentum, one will need to know its
self-energy as a function of \emph{energy} (i.e. the quantity that
should be extracted from the analysis). Similarly for an MDC, one
will need to know the matrix element as a function of
\emph{momentum}. Therefore, when the matrix element effect is
pronounced or the self-energy is strongly energy dependent,
fitting an MDC or EDC with a simple function (e.g. Lorenztian)
will not give a good agreement. Note that the implementation of
Kramers-Kronig transformation on self-energy can help to avoid
having assumption on the self-energy form and make it empirical
\cite{selfenergy:Norman, selfenergy:Chul}. However, such
transformation has limitation and requires the spectrum from
$-\infty$ to $+\infty$ in energy while clean data in doped
cuprates can only be obtained from Fermi level up to around half
of band width in binding energy before complications from valance
bands come in \cite{HEA:Non}.

As shown in Fig. 5, to compare the MDC and EDC peak dispersions,
we have constructed ARPES data (Fig. 5(b)) from the generated
Kramers-Kronig satisfied self-energy in Fig. 5(a) and the same
matrix element as used in Fig. 3(b). To generate this self-energy
in Fig. 5(a), we start out with our extracted self-energy of the
nodal spectrum (cut a) in Fig 2(m) and 2(n) and then we extend it
in an arbitrary but Kramers-Kronig consistant way. Given that all
information is known, the solution of poles ($\omega - \epsilon_k
- Re\Sigma(\omega)=0$) can be traced precisely as shown in Fig.
5(b). Although MDC and EDC peak dispersions show agreement for
small binding energy less than 0.3 eV, the discrepancy is large at
energy above the high-energy anomaly ($ > $ 0.3 eV.) In Fig. 5,
where the band is very dispersive, EDC analysis fails to track the
band since the EDC peak is not well-defined. On the other hand
(not shown), when the band is shallow (e.g. Fig. 2(c)), MDC peak
dispersion may fail to describe the band dispersion (discussed
also in Ref. \cite{HEA:Donglai}.)

\section{\label{other issues} Kramers-Kronig relation}
From Eq. 2, by causality, the real and imaginary part of
self-energy are related by Kramers-Kronig relations. In principle,
if the full spectral function ${\mathcal A}({\bf k},\omega)$ is
known, one could perform an inversion to obtain the full
self-energy using the Kramers-Kronig transformation
\cite{selfenergy:Norman, selfenergy:Chul}. However, such
transformation has limitations and requires the spectrum from
$-\infty$ to $+\infty$ in energy. Unfortunately, clean ARPES data
from doped cuprates can usually be obtained from Fermi level to
around half of the band width where complication of valance bands
will come in \cite{HEA:Non}.

Attempting to use the Kramers-Kronig transformation on self-energy
of doped-cuprate data is then required to have a cut-off/extension
model at energies above the existing data points
\cite{KK:Kordyuk}. However a cut-off/extension model is difficult
to be justified and the result can vary substantively, depending
on the cut-off/extension model used. For example, with a certain
assumption of cut-off/extension model, Ref. \cite{ME:Inosov}
claims that the self-energy extracted by using LDA as a bare band
cannot satisfy Kramer-Kronig relation. In contradiction to the
claim, here, by extending the self-energy in arbitrary form but
still having similar line shape as calculations from Ref.
\cite{Hubbard:Alexandru} or \cite{Magnon:Markiewicz}, Fig. 5(a)
shows that our self-energy extracted by using LDA could satisfy
Kramer-Kronig condition. In conclusion, given the limitation of
obtaining the whole band of doped cuprates, the implementation of
Kramer-Kronig relation is not possible without a further
assumption (e.g. cut-off/extension model) while it is found that
such assumption on cut-off/extension model can be highly sensitive
and difficult to be justified.

With the above reason, instead of attempting to implement
Kramers-Kronig condition in our analysis, we obtain self-energy by
using LDA as a reference of the bare band. By this way, in
principle, a self-energy obtained from 2D analysis will at least
be same from one to another ARPES measurement if referring to LDA
calculation which is a robust and mature technique.


\begin{thebibliography}{10}
\bibitem{TB:comment}
LDA calculation is represented approximately well by this set of
tight-binding parameters only in this fitting energy range, 0 -
0.7 eV where the band bottom of actual LDA calculation is slightly
larger than from these tight-binding parameters (please see Ref.
3).
\end{thebibliography}
\end{document}